# From Data Quality for AI to AI for Data Quality: A Systematic Review of Tools for AI-Augmented Data Quality Management in Data Warehouses

Heidi Carolina Tamm and Anastasija Nikiforova

**Abstract.** While high data quality (DQ) is critical for analytics, compliance, and AI performance, data quality management (DQM) remains a complex, resource-intensive and often manual process. This study investigates the extent to which existing tools support AI-augmented data quality management (DQM) in data warehouse environments. To this end, we conduct a systematic review of 151 DQ tools to evaluate their automation capabilities, particularly in detecting and recommending DQ rules in data warehouse - a key component of data ecosystems. Using a multi-phase screening process based on functionality, trialability, regulatory compliance (e.g., GDPR), and architectural compatibility with data warehouses, only 10 tools met the criteria for AI-augmented DQM. The analysis reveals that most tools emphasize data cleansing and preparation for AI, rather than leveraging AI to improve DQ itself. Although metadata- and ML-based rule detection techniques are present, features such as SQL-based rule specification, reconciliation logic, and explainability of AI-driven recommendations remain scarce. The study contributes practical guidance for tool selection and identifies critical design requirements for next-generation AI-driven DQ solutions—advocating a paradigm shift from "data quality for AI" to "AI for data quality management."

## 1 Introduction

In today's data-driven era, data serves as a critical asset, enabling the transformation of raw facts into actionable insights for decision-making across industries. However, the utility of these insights depends on data quality (DQ), a concept gaining attention since the 1960s and becoming prominent in computer science by the 1990s (Scannapieco & Catarci, 2002; Nikiforova, 2020). The projected growth of global data to 175 zettabytes by 2025 (Coughlin, 2018) amplifies the challenge of ensuring high-quality data while balancing storage and processing efficiency. Poor DQ carries significant costs - up to 19% of businesses report customer loss due to inaccurate or incomplete data (Dixon, 2020). The emergence of AI and Large Language Models (LLMs) further raises the stakes, as these systems depend on high-quality inputs to function effectively. This has led to widespread emphasis on "*data quality for AI*." However, leveraging AI to enhance DQ management itself, i.e., reversing the paradigm, remains underexplored.

Despite the ongoing shifts to decentralized, domain-driven architectures (Blohm et al., 2024), traditional data warehouses (DW) continue to serve as central infrastructures for integrating and analyzing organizational data (Blohm et al., 2024). These systems aggregate data from disparate sources across data ecosystem(s) (within and outside an organization's data ecosystems) but are often plagued by complex and time-consuming data quality management (DQM) tasks. Tracing data lineage and defining rules is particularly burdensome. Compliance obligations, such as GDPR, add to the cost and complexity of maintaining DQ (Karkošková, 2022).

To address these challenges, automation -and particularly AI-driven automation- holds strong promise. Considering relative predictability of DQ requirements within DWs environment (Fadler & Legner, 2020), we assume such solutions are already widely available in the market, seeking for the most appropriate for being adopted. This study, inspired by the challenges faced by a financial institution seeking to modernize its DQM (one of authors belongs to), investigates whether the market offers tools that support automated DQM, particularly rule detection and anomaly identification in DWs. From both a practitioner and research perspective, identifying such tools is a first step toward engineering more effective, AI-enhanced DQM systems.

We conduct a systematic review of 151 tools, assessing their functionality, integration with DWs, regulatory compliance, and support for rule discovery. Ultimately, only 10 tools met the defined criteria and were capable detecting DQ rules or anomalies automatically. Our findings suggest that while some tools offer ML or metadata-driven features, the landscape remains fragmented and lacks comprehensive AI augmentation, with current tools most often prioritizing data cleaning for AI applications, rather than using AI to improve DQ. As such, ML is rather seen as a "consumer" of DQ with the vast of research and developments on ensuring DQ for ML, with limited use of ML for DQM. Thus, we advocate a paradigm shift from the traditional focus on ensuring DQ for ML models to using AI and ML to improve data quality management, i.e., from "*Data Quality for AI*" to "*Data quality for AI and AI for Data Quality Management*".

The paper is structured as follows: Section 2 provides the background, Section 3 presents the methodology, Section 4 presents results, Section 5 discusses findings, limitations, and future directions, and the final section concludes the study.

## 2 Background

This section provides a brief overview of foundational concepts and a review of related literature to contextualize the study.

### 2.1 Concepts

Data quality refers to the extent to which data meets specific requirements, commonly aligned with ISO 9000[1] standards. Definitions range from abstract notions of "*fitness for use*" (Wang & Strong, 1996; Batini & Scannapieco, 2016) to measurable dimensions such as *completeness*, *timeliness*, *accuracy*, and *consistency* (Scannapieco & Catarci, 2002). These dimensions are context-dependent, varying by domain, data type, and use case and often evolving over time (Cichy & Rass, 2019; Nikiforova, 2020).

[1] https://www.iso.org/standard/62085.html

Efforts have been made to standardize DQ dimensions for specific sectors (Batini et al., 2009; Sidi et al., 2012), e.g., the European Parliament and Council mandate seven dimensions for financial institutions, which include *completeness*, *accuracy*, *consistency*, *timeliness*, *uniqueness*, *validity*, and *traceability* (Parliament & Council, 2013). However, these remain largely domain-specific and lack general applicability.

DQ assessment typically combines subjective -user evaluation- and objective -computational techniques- approaches (Batini et al., 2009; Lacagnina et al., 2023). Objective methods include detecting incorrect values, constraint violations, and integrity issues, highlighting the interplay between technical tools and stakeholder needs (Batini et al., 2009; Lacagnina et al., 2023). Effective DQM follows a top-down approach, translating business needs into enforceable rules categorized as *business DQ rules*, which describe quality expectations in business terms, and *data quality rule specifications*, which define physical-level requirements (Plotkin, 2020). Profiling and validation tasks help quantify DQ and ensure adherence to such rules (Loshin, 2010).

Data warehouses, integral to large organizations, consolidate historical and operational data across systems and act as a "*single source of truth*" (SAP, 2023), often, however, being issue prone. DQ issues may stem from poor input data (e.g., entry errors, database design flaws) or from integration and migration processes (Liu et al., 2019). These include formatting issues, missing records, and duplicates, often addressed through reconciliation and conformance checks (Experian, 2023). Metadata -physical, logical, and conceptual- plays a vital role in DQ, supporting traceability, integration, and rule generation (Hedden, 2016). As such, metadata is increasingly used as a foundation for automated DQ analysis methods.

### 2.2 Related Work

To inform this study, we conducted a systematic literature review (SLR) of existing literature on surveys of DQ tools, as well as the SLR on automated DQ rules detection. As this is not a central focus of this study, we do not provide methodological details about the conducted SLRs, which, however, are available on Zenodo [link will be added upon acceptance].

Our *SLR of DQ tools surveys* revealed three relevant studies: (Ehrlinger & Wöß, 2022), (Houston et al., 2018), and (Neely et al., 2006). Neely et al. (2006) evaluated tools in engineering asset management, and Houston et al. (2018) explored tools for clinical trials. Both studies primarily focused on domain-specific tools rather than on automated DQ rule detection. Given the rapid evolution of technology they are also now somewhat outdated today. Ehrlinger & Wöß (2022) reviewed 667 tools, identifying only 13 capable of automating routine tasks such as scheduling checks. However, none addressed automated DQ rule generation, central to this study's objectives.

Our *SLR on automated DQ rule generation* identified 10 relevant studies (available on Zenodo [link will be added upon acceptance]), with several more studies published in 2024 and 2025 (with SLR conducted in late 2023). Most studies focus on integrity constraints (ICs) rather than comprehensive DQ rules. These ICs include techniques for identifying data inconsistencies or constraints (Ilyas & Chu, 2015), with newer approaches extended to optimizing rule discovery for big data (Li et al., 2015; Taleb &

Serhani, 2017; Fan et al., 2022). Advanced approaches, such as study by Fan et al. (2022), include entity-enhancing rules, which combine ML and rule-based methods to address both entity and conflict resolution, and Heine et al.'s *RADAR* - a domain-specific language employing Autoregressive Integrated Moving Average (ARIMA) models for DQ rule specification (Heine et al., 2019). Recent advancements in automated DQ include *SAGED* - an error detection tool introduced by Abdelaal et al. (2024), which leverages few-shot meta-learning to detect errors in data being added to the system, generating feature vectors through meta-classifiers. Sartore et al. (2024), in turn, presented an anomaly detection system using fuzzy logic for agricultural data editing developed for the United States Department of Agriculture's National Agricultural Statistics Service. Finally, Ehrlinger et al. (2021) presented prototype called *DQ-MeeRKat* that offers automated monitoring through reference-data-profile-annotated knowledge graphs to verify that newly inserted or up-dated data continues to conform to the constraints stored in the reference-data-profiles, with the intent to come up in the future to achieve what the authors call "AI-based surveillance state", which would be capable of characterizing various kinds of data to detect drifts and anomalies in DQ at early stages.

Despite progress, most proposals emphasize specific dimensions such as uniqueness or domain-specific use cases, with the lack of solutions comprehensively detecting DQ rules, while specializing in data warehouses. Moreover, early stages of our SLR demonstrated that most DQM studies focus on ensuring DQ for ML tasks (Byabazaire et al., 2020; Li et al., 2024; Lu et al., 2023)), neglecting AI and ML's potential in augmenting DQM itself. This underscores the need for a paradigm shift and motivates the present study, which explores market tools capable of addressing these limitations.

# 3 Methodology

This study employs a systematic review methodology (Kitchenham & Brereton, 2013), adapted to examine DQ tools. The objective is to identify tools that leverage AI -ML or alternative methods- for automatically detecting DQ rules and anomalies, while also allowing users to define custom rules for DQ adjustments. To attain this objective, five key **research questions** were established.

***Q1***. *What is the current landscape of DQ tools?*

***Q2***. *What functionalities do DQ tools offer?*

***Q3***. *Which data storage systems DQ tools support? and where does the processing of the organization's data occur?*

***Q4***. *What methods do DQ tools use for rule detection?*

***Q5***. *What are the advantages and disadvantages of existing solutions?*

To address these questions, tools were identified through a combination of rankings from technology reviewers and academic sources. A Google search was conducted using keyword *("the best data quality tools" OR "the best data quality software" OR "top data quality tools" OR "top data quality software") AND "2023"* (search conducted in December 2023). Additionally, this list was complemented by DQ tools found in academic articles, identified with two queries in Scopus, namely *"data quality tool"*

*OR "data quality software"* and *("information quality" OR "data quality") AND ("software" OR "tool" OR "application") AND "data quality rule".*

For **selecting** DQ tools, several exclusion criteria were applied. Tools from sponsored, outdated (pre-2023), non-English, or non-technical sources were excluded. Academic papers were restricted to those published within the last ten years, focusing on the computer science field. This resulted in 16 ranking lists[2], and 35 academic papers, and 151 DQ tools. A list of sources and tools is provided on Zenodo [link will be added].

To structure the **review** process and facilitate answering the established questions (Q1-Q3), a review protocol was developed, consisting of three sections (protocol is available on Zenodo – [link to be added upon acceptance]). The initial tool assessment was based on *availability*, *functionality*, and *trialability* (e.g., open-source, demo version, or free trial). Tools that were discontinued or lacked sufficient information were excluded. The second phase (and protocol section) focused on evaluating the functionalities of the identified tools. Initially, the core ***DQM functionalities*** were assessed, such as *data profiling, custom DQ rule creation, anomaly detection, data cleansing*, *report generation*, *rule detection, data enrichment*. Subsequently, *additional* ***data management functionalities*** such as *master data management, data lineage, data cataloging, semantic discovery*, and *integration* were considered. These parameters were selected based on the authors' experience with DQ tasks, discussions with DQ professionals, and previous research (Ehrlinger & Wöß, 2022).

The final stage of the review examined the tools' *compatibility with data warehouses* and General Data Protection Regulation (GDPR) compliance. Tools that did not meet these criteria were excluded. As such, the 3rd section of the protocol evaluated the tool's environment and connectivity features, such as whether it operates in the *cloud*, *hybrid*, or *on-premises*, its *API support*, *input data types (.txt, .csv, .xlsx, .json)*, and its ability to connect to data sources including *relational* and *non-relational databases*, *data warehouses*, *cloud data storages*, *data lakes*. Additionally, it assessed whether the tool processes data *on-premises* or in the *vendor's cloud environment*. Tools were excluded based on criteria such as not supporting data warehouses or processing data externally.

As such, the review applied the following selection criteria: (1) Exclusion Criteria (EC): (1a) *EC1: Tool does not exist*; (1b) *EC2: Tool is outdated or has been discontinued*; (1c) *EC3: Tool does not qualifies as a DQ tool*; (1d) *EC4: Tool is part of another tool, integrated system or suite*; (1e) *EC5: Insufficient information available about the tool*; (1f) *EC6: Tool only checks only a single data attribute*; (1g) *EC7: Tool does not detect DQ rules or anomalies*; (1h) *EC8: Tool detects anomalies, but does not support the definition of DQ rules*; (1i) *EC9: Tool is not suitable for data warehouses*; (1j) *EC10: Data processing location is unclear*; (1k) *EC11: Tool processes data exclusively in the vendor's cloud environment*; (2) Inclusion Criteria (IC): (2a) *IC1: Tool supports automated DQ rule detection*; (2b) *IC2: Tool is capable of detecting anomalies and allows users to define custom DQ rules*.

[2] Datamation, Simplilearn, TechTarget, Solutions Review, TechRepublic, Geekflare, TrustRadius, BIS (Grooper), G2, Slashdot, SourceForge, PeerSpot, SoftwareReviews, WebinarCare, HubSpot, Gartner

Finally, DQ tools were reviewed, and data **synthesized.** Tools were reviewed based on testing of tools, information on official websites, demos, and documentation.

## 4 Results and Analysis

### 4.1 The Data Quality Tools Landscape

RQ1 examined the availability and characteristics of DQ tools, focusing on their existence, functionality, and availability for testing or demos. Initially, tools were filtered using the EC1-EC5 exclusion criteria. We evaluated the availability of trials and documentation and assessed the level of detail in the information provided for each tool. A detailed summary of analysis results is available on Zenodo [link to be added].

The search identified 151 DQ tools, with 46 excluded during validation (Fig. 1). Four tools were found to *no longer exist* (*Data Preparator, DataMentors, Synchronos*, and *matchIT DQ Solutions*), six were *discontinued or marked legacy* (e.g., *Datiris Profiler* and *Experian Pandora*), and nine tools were "*duplicates,*" or *"integrated into larger platforms"* (e.g., *Rapid Data Profiling* and *Self-Service Data Preparation* within *DataRobot AI Platform*). Additionally, 27 tools were found to be *irrelevant to DQ*, primarily focusing on functionalities such as data integration or customer management. After exclusions, 105 tools remained for further analysis.

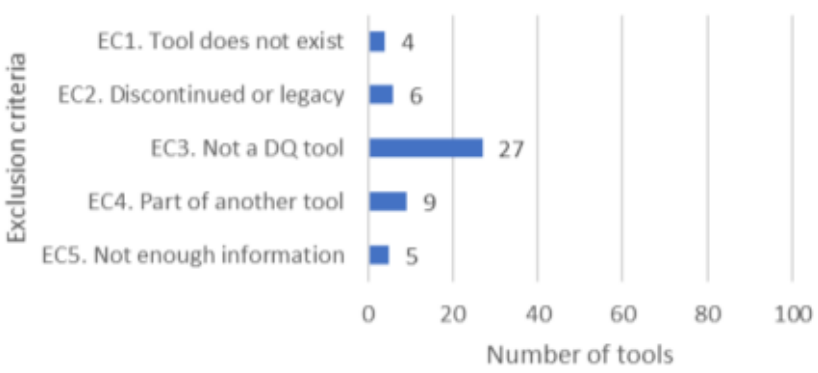

**Fig** 1.Tools excluded at the 1st phase.

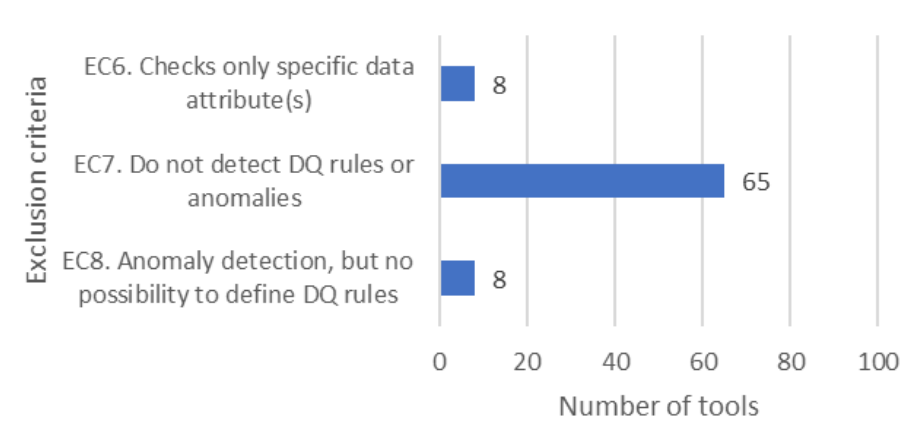

**Fig.** 2.Tools excluded at the 2nd phase.

Trial availability varied: 13 tools were *open-source*, 10 offered *free trials*, 5 provided *demos*. 12 tools had a trial request form, and 43 offered demo requests, though most received *no response*, with some resulted in *sales calls*, with 2 cases resulting in a *software provider call*, and one - *free license* for a trial. A total of 22 tools were *non-trialable*, requiring direct purchase. Despite these limitations, no tools were excluded based solely on trial availability. For non-trialable tools, reviews were based on available documentation, websites, and videos. Around half provided public documentation, while 45 (42.9%) did not. Some tools lacked sufficient details, providing heavily generic marketing content. To address this, we introduced a "*Level of Information*" attribute to assess the clarity and completeness of the provided materials. Of the reviewed tools, 72 were well-described, 28 had partial descriptions, and 5 were excluded due to insufficient information. Several tools lacked clear functional descriptions. For instance, the *Black Tiger Platform* provided vague details about its DQ features, and tools such as *DataStreams* and *OpenDQ* were unclear on their DQ capabilities beyond validation, reporting. *Deduplix Ixsight* mentioned fuzzy matching models, but their functionality was not explained in detail. In the first phase, 51 tools were excluded, leaving 100 tools for further analysis, as shown in Fig. 1.

### 4.2 Features of Data Quality Tools

Next, we examined the features and functionality of DQ tools, particularly their ability to automatically detect DQ rules. Detailed results are available on Zenodo [Link to be added]. The 100 tools were mapped to the DQ functionalities (Fig. 3). The most popular DQ functionalities included *data cleansing* (75%) and *profiling* (67%) were the most common features, while only 12% of tools supported *DQ rule detection*. *SQL-based rule definition*, critical for data warehouse users, was the least common, present in 6% of tools.

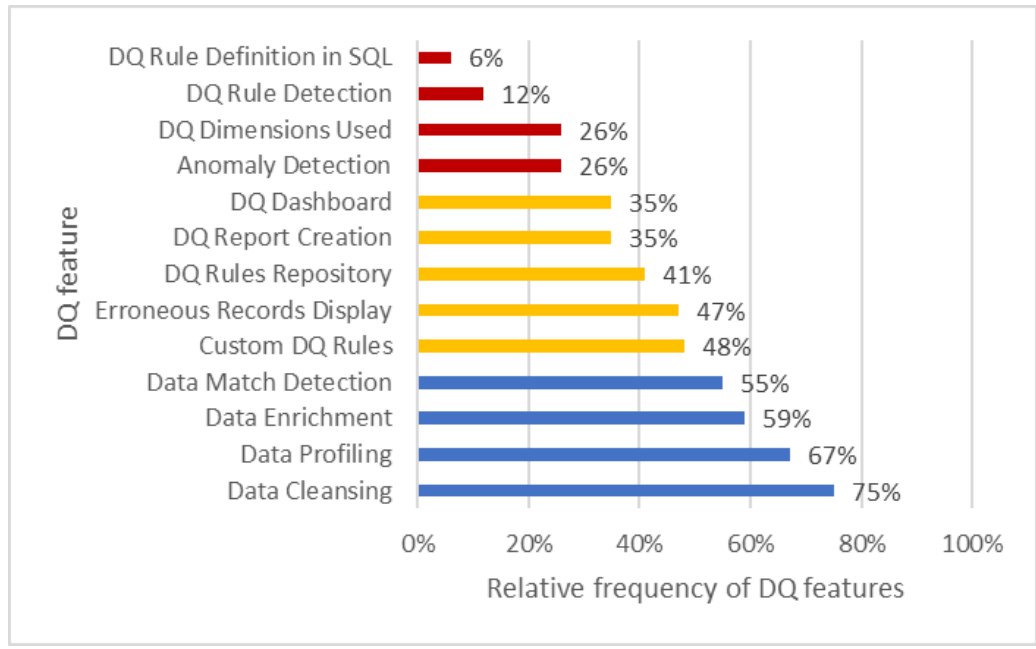

**Fig. 3.** Relative frequency of features

The tools were also evaluated for *additional functionalities* such as *master data management*, *data lineage*, *cataloguing*, *semantic discovery*, and *data integration,* with their relative frequencies presented in Fig. 4. Some tools appeared to be purely DQM-focused (e.g., *OpenRefine*, *Ataccama DQ Analyzer*), while others were multi-functional platforms (e.g., *SAP Information Steward, Syniti Knowledge Platform*).

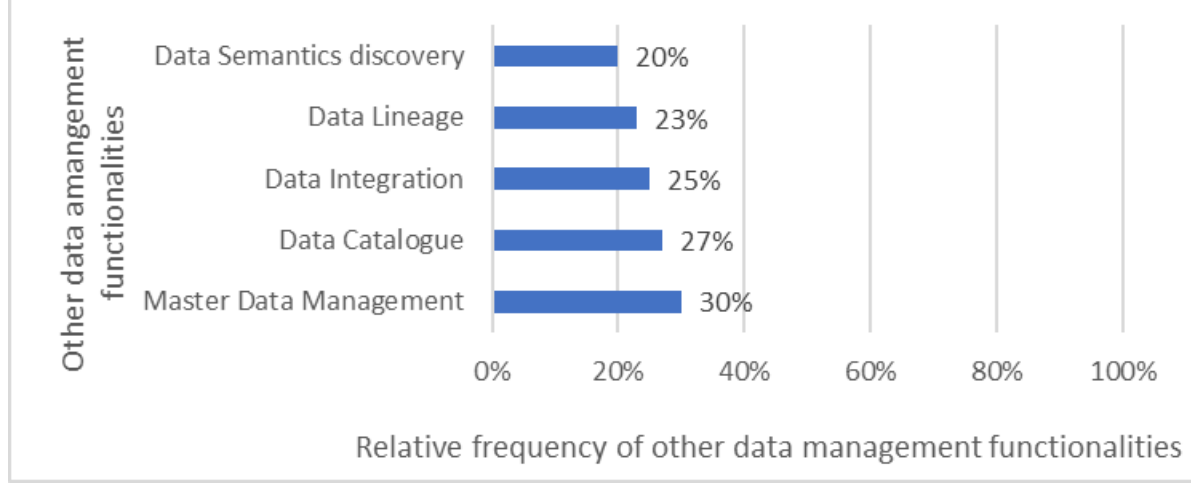

**Fig. 4.** Relative frequency of other data management functionalities.

Eight tools were narrowly focused on specific attributes (e.g., *email*, *address validation*) or anomaly detection but lacked custom rule definitions (incl. based on the discovered anomalies). Examples include *Experian Email Validation*, *Informatica Address Verification, Holodetect*, *Rapid Data Profiling,* and *Talend Data Fabric*. 65 tools lacked both *anomaly detection* and *DQ rule detection*, with 54 being cleansing-focused, with 31 of them lacking *custom rule definition* (e.g., *TIBCO Clarity*, *OpenRefine*). These tools are primarily suited for data preparation for ML tasks. After applying EC6-EC8 exclusion criteria, 19 tools remained - 12 that could detect DQ rules and 7 that detected anomalies and allowed custom rule definitions.

The final 19 tools were further analysed for their ability to detect DQ rules or anomalies, define custom DQ rules, and other functionalities (Table 1). Tools capable of detecting and recommending DQ rules (IC1) were more multifunctional than those under IC2[3], i.e., were more likely to allow custom rule definitions. For IC1 tools, SQL rule definition was rare (8.3%). Data management functionalities (Table 2) were more frequent among IC1 tools, further supporting their multifunctional nature.

**Table 1.** Relative frequency of DQ features.

| Functionality | IC1 | IC2 |
|---|---|---|
| Custom DQ Rules | 100% | 100% |
| DQ Rules Repository | 91.7% | 100% |
| Anomaly Detection | 91.7% | 100% |
| Data Profiling | 100% | 85.7% |
| Erroneous Records Display | 100% | 71.4% |
| DQ Report Creation | 91.7% | 71.4% |
| DQ Dashboard | 75% | 85.7% |
| DQ Dimensions Used | 75% | 57.1% |
| Data Match Detection | 75% | 42.9% |
| Data Cleansing | 75% | 42.9% |
| DQ Rule Detection | 100% | 0% |
| Data Enrichment | 50% | 28.6% |
| DQ Rule Definition in SQL | 8.3% | 57.1% |
| Custom DQ Rules | 100% | 100% |

**Table 2.** Relative frequency of other data management functionalities.

| Functionality | IC1 | IC2 |
|---|---|---|
| Data Semantics discovery | 75% | 57.1% |
| Data Catalogue | 75% | 57.1% |
| Data Lineage | 75% | 42.9% |
| Master Data Management | 66.7% | 14.3% |
| Data Integration | 41.7% | 28.6% |

The included tools (IC1 and IC2) offered broader functionality than excluded tools (Fig. 5), which primarily focused on data cleansing and enrichment. This aligns with the study's aim to identify tools for DQ issue detection rather than issue fixing, as data warehouses rely on correcting issues at the source before loading data.

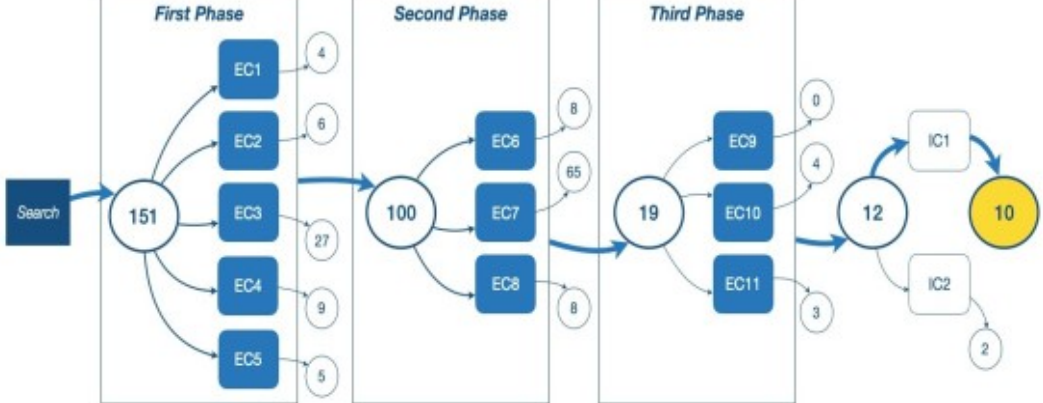

**Fig. 5.** Review process.

[3] Data enrichment and cleansing functions are mapped for statistical purposes but are not in the scope of this study as it aims to look for DQ tools for data warehouses where data cleansing and enrichment are not used locally.

### 4.3 The Environment and Connectivity

This section evaluates the environment solutions and connectivity features (RQ3) of the 19 DQ tools. Detailed results are available on Zenodo [Link to be added upon acceptance]. All 19 tools supported *connections to data warehouses*, incl. compatibility with relational and non-relational databases, cloud data storages, data lakes, and popular systems such as *Teradata Vantage, Snowflake, and Amazon Redshift*. As such, no tools were excluded. Then, tools were analyzed with relation to their *deployment environments*. 13 tools were *cloud-based*, two tools operated both *on-premises* and in the cloud, one - on-premises, one - hybrid, and one tool offered cloud or *hybrid deployment*.

Regarding *data processing location*, 7 tools supported processing on *the vendor's or organization's side*, 4 processed data in *private clouds*, one - *on-premises*, three - in the *vendor's cloud*, and four - *lacked information* on processing locations. To ensure GDPR compliance, tools with unknown processing locations or vendor-cloud-only processing were excluded, leaving 12 tools - 10 capable of detecting DQ rules, and 2 alternatives (IC2) that detected anomalies and allowed custom rules definition.

Overall, the review process involved three phases, narrowing down from 151 tools to 12 suitable candidates ((based on EC1-EC11), Fig. 6). Initial *validation* (EC1-EC5) excluded tools that did not exist, were discontinued, or lacked sufficient information. This reduced the list to 100 tools with DQ functionalities. Exclusions based on *inability to detect DQ rules or anomalies, along with the custom rule definition* (EC6-EC8) reduced the list to 19 tools - 12 DQ **rule detectors** (IC1) and 7 **anomaly detectors** with custom rule definition (IC2). Tools *failing to meet environmental and connectivity expectations* (EC9-EC11) were excluded, leaving 12 tools - 10 meeting the main goal (IC1) and 2 alternatives meeting IC2. These are *AbInitio Enterprise Data Platform, Anomalo, Ataccama ONE, Collibra Platform, DQLABS Platform, DvSum, Global IDs Data Quality Suites, Informatica Cloud Data Quality, Informatica Data Engineering Quality, Informatica Master Data Management, LiTech Data Quality Management*.

Among the non-trialable tools, some allowed demo requests or provided documentation. Open-source tools were significantly reduced, with 11 excluded as lacking DQ rule detection (EC7) or custom rule definitions (EC8). All tools with available free trials were also excluded for the same reason.

Some alternative DQ tools were identified as semi-automated DQ rule detection solutions being able to detect anomalies and enable users to define their own DQ rules. Among the 7 anomaly detection tools, for 4 tools, it was unclear where the data is processed, and one tool processed data on the vendor's cloud. Following the exclusion criteria, only 2 suitable solutions remained: (1) *Anomalo* that uses unsupervised ML to detect DQ issues without predefined rules or thresholds, allowing users to adjust monitoring without coding; (2) *LiTech DQ Management* that consolidates data validations into a single platform, encompassing a DQ rule repository and DQ reports. It leverages ML to autonomously generate DQ validations, including anomaly detection with an integrated alerting system.

Other anomaly detectors (IC2) primarily focused on data preparation for ML or business analytics, offering functionalities such as *data cleansing* and *enrichment* but with-out using DQ rules, with examples including *Experian DataArc360, Rapid Data Profiling*. The heatmap in Fig. 6 visualizes tool selection based on defined criteria, emphasizing the challenges in identifying suitable tools for AI-augmented DQM in DW.

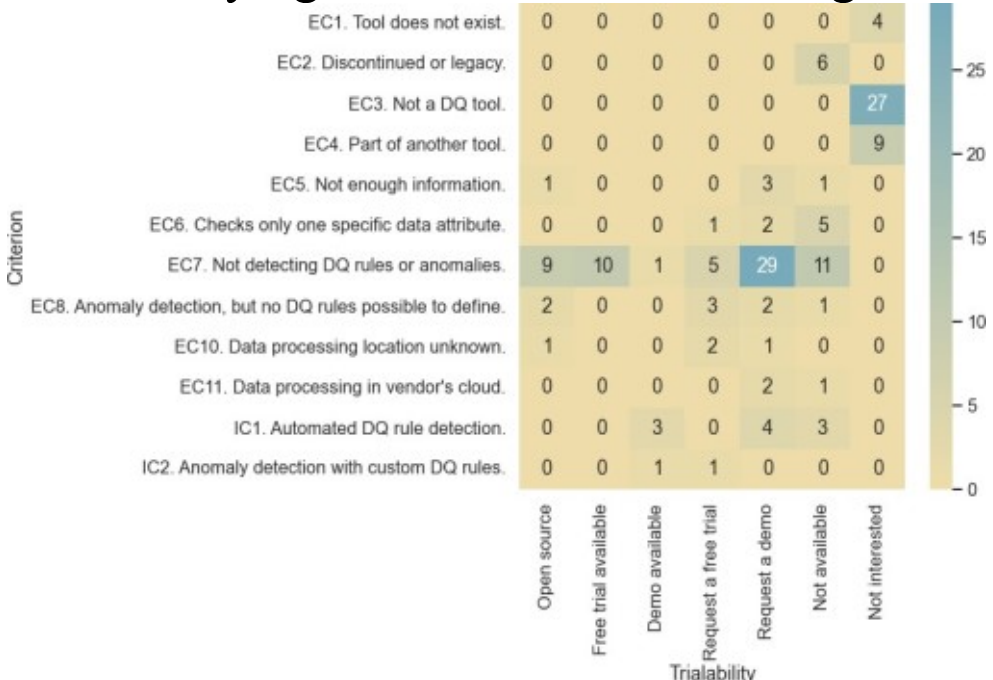

**Fig. 6.** Heatmap of tool counts by selection criteria and trialability.

### 4.4 Solutions supporting the Data Quality Rule Detection

As a result, only ten DQ tools satisfied the defined criteria, being able to detect DQ rules and suited for use with data warehouses. Our analysis identified four primary methods (RQ4) employed by these tools for DQ rules discovery (detailed tool-by-tool description is available on Zenodo [Link to be added upon acceptance]): (1) **metadata-based detection** (*DQLabs Platform*); (2) **built-in rules and ML** (*Ataccama ONE Platform*, *DvSum*); (3) **metadata and ML** (*AbInitio Enterprise Data Platform*, *Informatica* products); (4) **ML-only detection** (*Collibra*, *Syniti Knowledge Platform*).

Five tools emphasize *metadata* as the foundation for rule discovery, while six incorporate ML. One tool, *Global IDs DEEP Platform*, provides limited details on its approach to DQ rule detection, however, emphasizes *metadata management* and *data lineage*. As such, it can be inferred that metadata serves as an essential foundation for (AI/)ML-driven DQ rules discovery. However, solutions tailored specifically to data warehouses and their complex ecosystems remain scarce.

Finally, all ten tools were cloud-based and connected to data sources via APIs, ensuring broad compatibility and flexibility across various data environments, including public, private, and virtual private clouds.

### 4.5 Advantages and Disadvantages of Current Solution

As part of RQ5, we examined the strengths and limitations of the DQ tools capable of detecting DQ rules, to derive insights for future advancements. Key features of these tools are summarized in Fig. 7, with detailed breakdown of features per tool available on Zenodo. Our analysis revealed that all the 10 DQ tools form the final pool possess the capability for *data profiling*, enabling users to define *custom rules*, *generate DQ reports*, *maintain and organize rules in DQ rule repository*, and *report erroneous records for their further investigation*, and the ability to connect to data warehouses via APIs. Some tools support management of *DQ rules* by allowing them to be *edited, accepted, and rejected*, along with *tagging DQ rules with relevant data elements and* *business terms*. As such, they have been categorized as advantages and “minimum requirements” for AI-augmented DQM.

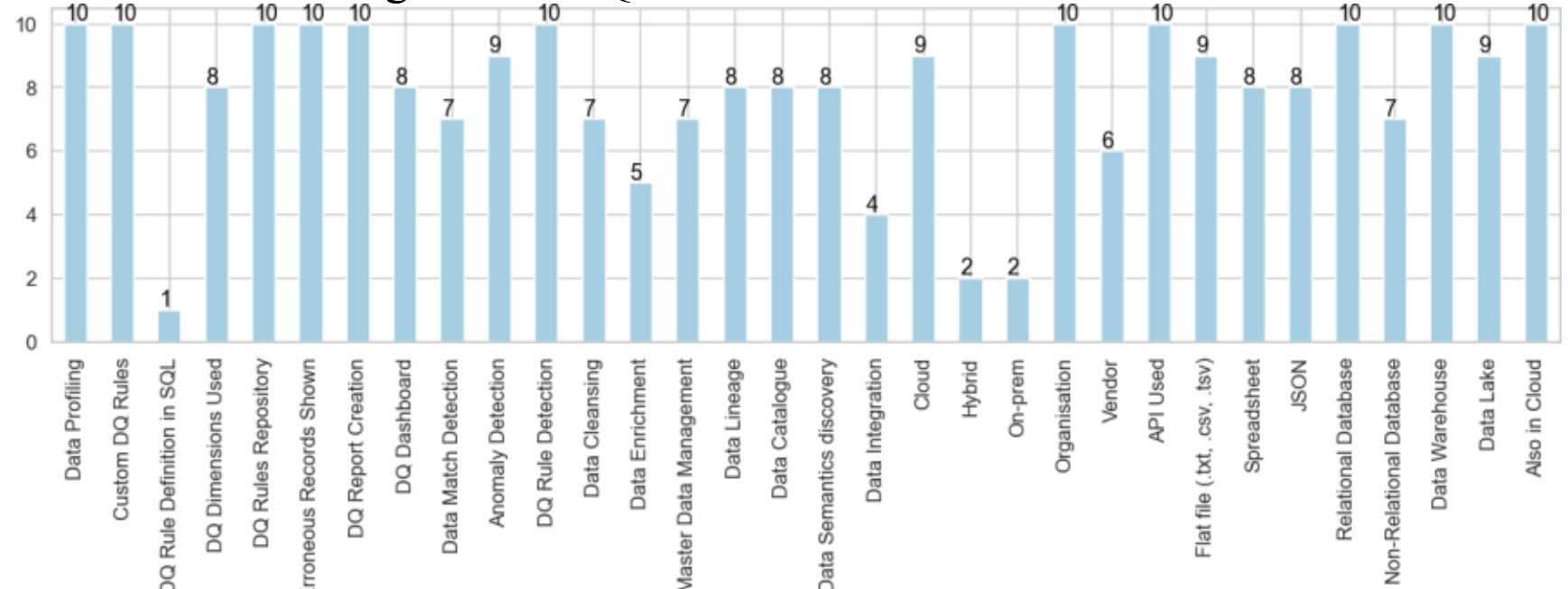

**Fig. 7.** Frequencies of DQ, other data management, environment, and connectivity features.

While these tools offered valuable functionality, several gaps were identified: (1) lack of support of the *detection of reconciliation rules*, critical for ensuring data consistency across systems; (2) *SQL-based rule definition* was rare, despite its relevance for data warehouse users; (3-4) many tools lacked *support for diverse data types* (e.g., integer, float, Boolean, string, date), with many tools rather focusing on selected data types, ignoring others, *and failing to cover a broad range of DQ dimensions* (e.g., referential integrity, external consistency) or tag detected rules with relevant dimensions; (5) analysed tools often lacked transparency in rule recommendation logic, hindering stakeholder involvement to *modify, accept, or reject suggested rules* before implementation; (6) while *cloud computing* offers benefits such as scalability and flexibility, for many tools it remained unclear if existing solutions were maximizing these advantages.

Although current DQ tools demonstrate strengths, they lack critical capabilities, preventing the identification of a "silver bullet" solution for comprehensive AI-augmented DQM in data warehouses. These gaps highlight the need for further innovation to better serve data warehouse environments.

# 5 Discussion

## 5.1 Discussion and Implications

This study aimed to assess the extent to which current DQ tools support AI-augmented DQM in DW environments. Through a systematic review of 151 tools, we identified only 10 that met key criteria, including support for rule detection, metadata utilization, DW integration, GDPR-compliant deployment, and user-driven configurability. Such a drastic reduction aligns with prior studies, such as Ehrlinger & Wöß (2022), who found only 17 suitable tools out of 667, noting limitations such as proprietary focus or minimal DQ measurement features. The main barriers include discontinued products, limited functionality, and inadequate documentation or trial access, highlight the market’s fragmentation, making tool selection difficult for organizations.

The findings reveal a significant gap in both commercial and academic tool offerings. While many tools support data profiling and cleansing, few implement automation capabilities that actively leverage AI or ML to improve DQ (Byabazaire et al., 2020; Li et al., 2024). I.e., AI is largely treated as a *consumer* of high-quality data rather than an *enabler* of data quality processes.

Only a small fraction of tools utilized ML-based techniques, and those that did often lack transparency and user control. Metadata-driven methods are common but remained underutilized in dynamic rule generation. Few tools offered hybrid approaches that combine metadata inference, ML-based anomaly detection, and rule-based logic, even though such combinations show high potential (Fan et al., 2022; Heine et al., 2019).

Moreover, support for SQL-based rule definition essential in DW contexts was rare, despite being widely acknowledged as a critical feature (Plotkin, 2020; Cichy & Rass, 2019). Most tools also lacked support for tagging rules with DQ dimensions (Batini & Scannapieco, 2016), reconciliation logic, and collaborative rule validation between business and technical stakeholders.

As such, this study has several practical and theoretical implications. The findings reveal critical gaps in current tools for augmented DQM tailored to DW. This gap presents opportunities for the development and commercialization of advanced AI-augmented solutions that can detect and enforce DQ rules automatically, streamlining DQM processes, reducing human workload, and lowering operational costs while ensuring regulatory compliance with regulations such as GDPR. More specifically, our analysis, including of strengths and weaknesses of the analysed tools (Section 4.5), suggest key priorities for developing AI-augmented DQM tools: (1-2) **automated rule generation** enabling rule generation for diverse data types and domains, using metadata as the foundation for ML-driven rule discovery and anomaly detection, and leveraging Natural Language Processing (NLP) and LLMs to create **rules in both SQL and natural language formats** for technical and business stakeholders and their collaboration; (3) **comprehensive data types coverage** and domains (finance, healthcare, education, etc.), automatically determining the type based on the data to cover all data elements of DW, (4) **broad DQ dimension coverage** addressing commonly used dimensions and integrating domain-specific requirements, such as financial reporting standards (Parliament & Council, 2013) to be followed by financial institutions to report DQ (by specific set of DQ dimensions); (5) **efficient processing using cloud computing** for scalability, while (6) ensuring **in-stack or private cloud data processing for GDPR (and other regulation) compliance**; (7) **enhanced governance and collaboration by linking rules to roles** (e.g., data (quality) stewards, business analysts, information owners) to streamline governance and reduce rule overload. To move the field forward, future tools must address both technical and organizational needs, bridging the gap between automation and explainability, and between AI engineers, DQ stewards and business users.

For practitioners, the study offers actionable guidance for evaluating and selecting appropriate DQ tools based on functionality, explainability, deployment flexibility, and integration with existing DW infrastructures. With only 10% of widely promoted tools meeting criteria for DQM in DW, organizations can use this study to filter irrelevant tools and focus on testing viable candidates, improving tool selection and adoption.

The study advances the dialogue on leveraging AI, and ML, in particular, for automating (or at least augmenting) DQM. The study lays the groundwork for a theoretical framework for automated DQ rule detection (in DW), which so far has been limited to a very few studies, such as a ML-based solution for DQ controls as an essential instrument in DQM (Walter et al., 2022).

Moreover, this study challenges the dominant narrative of "*data quality for ML*" by advocating a shift to "***ML (or broader – AI) for data quality management***" advocating AI-driven tools that enhance DQ processes while ensuring high-quality data across ecosystems for analytics, ML, and informed decision-making.

The rise of LLMs suggests it is a perfect time to launch a dialogue on how emerging technologies, such as LLMs and Explainable AI (XAI), can revolutionize DQM and broader data management practices within data ecosystems (Pucci et al., 2024). LLMs presents an opportunity to integrate natural language generation and understanding into DQM processes, enabling more intuitive interactions and innovative functionalities, bridging the gap between technical data stewards and business stakeholders. They could enhance interpretability of detected DQ issues by providing detailed, context-aware explanations and recommendations in plain language. LLMs are also seen to have the potential to support metadata enrichment, improve semantic discovery, and enable more dynamic rule generation tailored to specific organizational needs (Fernandez et al., 2023; Kliimask and Nikiforova, 2024; Pernici et al., 2024; Varma et al., 2024). Despite their promise, the integration of LLMs into DQM is underexplored in academic literature, with most progress occurring in proprietary environments. This also aligns with Zhou et al. (2024), whose analysis–centred around DQ for ML though, identified 4 tools– *Winpure*, *Ataccama ONE*, *Soda Core*, and *Evidently*– that have stepped out to integrate AI, with *Soda Core* integrating generative pre-trained transformers (GPT) into the modelling, rules suggestion, and monitoring tasks, which makes DQ checks more friendly to non-technical users, such as product managers and business owners. Future research should prioritize open collaboration between academia, industry, and practitioners to harness LLMs and XAI for DQM to enhance data ecosystems.

## 5.2 Limitations

This study has several limitations. A primary limitation of this study is the reliance on secondary data for tools that were not available for testing, predominantly sourcing information from software providers' websites, videos, and documentation. This reliance may introduce bias or restrict comprehensive evaluation. Secondly, while the systematic search was thorough, there remains a risk of omitting some tools. Only 10 tools capable of automated DQ rule generation were identified from 151 initially considered, potentially narrowing the range of perspectives and innovations captured, although they were selected from over 50 sources. In addition, most tools were found to target data cleansing for domain-specific databases rather than addressing the unique needs of DW, limiting the generalizability of findings to DW context. Finally, the rapid evolution of the field, particularly with advancements such Generative AI, means the study represents a snapshot in time. Future research will be necessary to reflect ongoing developments. Nonetheless, this study provides insights into the current state and calls for further advancements tailored to data warehouses and/or emerging technologies.

# 6 Conclusion

This study systematically reviewed 151 DQ tools to evaluate their suitability for AI-augmented DQM in DW. Through a multi-phase evaluation process, only 10 tools were found to support automated rule detection and align with key organizational requirements, revealing a significant gap in both market and academic offerings. Core limitations include the lack of SQL-based rule definition, weak reconciliation support, limited transparency of AI-generated rules, and narrow data type or DQ dimension coverage. While metadata- and ML-based methods show promise, hybrid approaches combining metadata, ML, and rule-based logic, though considered most effective, remain rare.

The results reveal that most tools focus on ensuring "*data quality for AI*" rather leveraging AI for DQ management itself. AI is still largely treated as a consumer of data rather than an enabler of data quality processes. As AI and low-code technologies advance (Sundberg & Holmström, 2023), we anticipate the rise of user-friendly, self-service DQ tools, empowering non-technical users to conduct DQ evaluation and improvement tasks more independently and efficiently (Zhou et al., 2024).

As such, this study advocates renewed focus on integrating AI/ML into DQM to enable dynamic, explainable rule detection across data types and business domains. Key design requirements for future tools include SQL-based and natural language rule definition, explainable AI for business-technical collaboration, GDPR-compliant deployment, and scalable cloud architectures vital to enabling robust and accessible DQM within data warehouse ecosystems. The findings will inform the next phase of our research - requirements engineering for the development of an AI-augmented DQM tool, grounded in the strengths and limitations identified among the final pool.

Practically, the study offers a roadmap for organizations to evaluate and select suitable DQ tools aligned with DW-specific needs. Theoretically, it enriches the theoretical foundation for AI-driven DQM solutions in DW and beyond. It underscores the need for interdisciplinary collaboration to address the shortcomings of current tools, making a call for both helixes to shift the dominant narrative of "*data quality for ML*" toward ***"data quality for AI and AI for data quality management*,"** highlighting the potential of emerging technologies such as LLMs to transform DQM and data governance. This shift can empower organizations to harness the full potential of AI, building robust data ecosystems and maintaining them, ensuring high-quality data for diverse applications, and fostering progress in data management practices.

**Declaration of Generative AI and AI-assisted technologies in the writing process.** The authors hereby discloses that ChtGPT-3.5 was used to improve the conciseness and clarity of selected sentences in this study. After using this tool/service, the authors reviewed and edited the content as needed and take full responsibility for the content of the published article.

**Disclosure of Interests.** The authors have no competing interests to declare that are relevant to the content of this article.